\documentclass[preprint,11pt]{elsarticle}

\usepackage{amsmath,amscd}
\usepackage{amssymb}
\usepackage{cases}
\usepackage{ulem}
\usepackage{color}
\usepackage{tabularx}

\newtheorem{thm}{Theorem}
\newtheorem{lem}[thm]{Lemma}
\newtheorem{prop}[thm]{Proposition}

\newtheorem{rem}[thm]{Remark}
\newproof{pf}{Proof}

\newcommand{\bF}{ {\mathbb F}}
\newcommand{\EOP} { \hfill $\Box$ }

\begin{document}

\begin{frontmatter}



\title{Two types of permutation polynomials \\
with special forms }


\author[1]{Dabin Zheng}
\ead{dzheng@hubu.edu.cn}
\author[1]{Mu Yuan}
\ead{yuanmu847566@outlook.com}
\author[2]{Long Yu}
\ead{longyu@mails.ccnu.edu.cn}


\address[1]{Hubei Province Key Laboratory of Applied Mathematics, \\
Faculty of Mathematics and Statistics, Hubei University, Wuhan 430062, China }
\address[2]{School of Mathematics and Physics, Hubei Polytechnic University,\\ Huangshi 435003, China}

\begin{abstract}
Let $q$ be a power of a prime and $\bF_q$ be a finite field with $q$ elements. In this paper, we propose four families of infinite classes
of permutation trinomials having the form $cx-x^s + x^{qs}$ over $\bF_{q^2}$, and investigate the relationship between this type of permutation
polynomials with that of the form $(x^q-x+\delta)^s+cx$. Based on this relation, many classes of permutation trinomials
having the form $(x^q-x+\delta)^s+cx$ without restriction on $\delta$ over $\bF_{q^2}$ are derived from known permutation trinomials having the form $cx-x^s + x^{qs}$.
\end{abstract}

\begin{keyword}
Finite field, Permutation polynomial, Symbolic computation


\end{keyword}

\end{frontmatter}



\section{Introduction}

Let $q$ be a power of a prime. Let $\bF_{q}$ be a finite field with $q$ elements and $\bF_q^*$ denote its multiplicative group.
A polynomial $f(x)\in \bF_{q}[x]$ is called a {\it permutation polynomial} (PP) if its associated polynomial mapping $f: c\mapsto f(c)$
from $\bF_q$ to itself is a bijection. Permutation polynomials over finite fields have important applications in cryptography, coding
theory and combinatorial design theory. So, finding new permutation polynomials is of great interest in both theoretical and applied aspects.
The study of permutation polynomials has a long history~\cite{Dickson1896,Hermite1863} and many recent results are surveyed in~\cite{Hou2015}.

Permutation polynomials with few terms attracts many researchers' attention due to their simple algebraic representation
and wide applications in coding theory, combinatorial designs and cryptography. The recent progress on construction of permutation binomials and trinomials
can be seen in~\cite{BassalygoZinoviev2015,DingQuWangYuanYuan2015,GuptaSharma2016,LiHelleseth2016,LiQuWang2017,LiHelleseth12016,LiHellesethTang2013,LiQuChen2017,MaGe2017,TuZengHu2014,ZhaHuFan2017}
and the references therein. Permutation polynomials of the form $x +\gamma{\rm Tr}_{q}^{q^n}(x^k)$ were investigated in~\cite{CharpinKyureghyan2008,CharpinKyureghyan2010}
for some special $\gamma\in \bF_{q^n}$. Very recently, G.Kyureghyan and M.E.\ Zieve \cite{KyureghyanZieve2016} found almost all permutation polynomials over $\bF_{q^n}$  of
this form for $\gamma\in \bF_{q^n}^*$ and $q^n\leq 5000$, where $q$ is odd. And Li et al.~\cite{LiQuWang2017} studied permutation behavior
of this type of polynomials over the finite fields with even characteristic. In both~\cite{KyureghyanZieve2016} and~\cite{LiQuWang2017}, most permutation polynomials
having the form $x +\gamma{\rm Tr}_{q}^{q^n}(x^k)$ occur in case of $n=2$, i.e., they are trinomial permutations. Inspired by this, we construct four families of
infinite classes of permutation trinomials over $\bF_{q^2}$ with the following form
\begin{equation}\label{eq:ojpoly1}
cx -x^s + x^{qs},
\end{equation}
where $s$ is a positive integer and $c\in \bF_{q^2}$. We use the well-known result (Lemma~\ref{lem:PLWZ}) provided by Park, Lee, Wang and Zieve
to prove the first three classes of permutation trinomials. For the proof of permutation behavior of the last class of trinomials,
symbolic computation method related to Gr$\ddot{o}$bner bases and resultants is used.

On the other hand, in order to derive new Kloosterman sums identities over $\bF_{q^n}$,
Helleseth and Zinoviev~\cite{HellesethZinoviev2003} found a class of permutation polynomials related to the form
\begin{equation}\label{eq:ojpoly2}
(x^{q^i}-x+\delta)^s +cx
\end{equation}
over $\bF_{q^n}$, and Ding and Yuan~\cite{DingYuan2007} first studied this type of permutation polynomials.
Following line of work in~\cite{DingYuan2007}, many permutation polynomials with this form have been constructed~\cite{ CepakCharpinPasalic2017,LiHellesethTang2013,TuZengJiang2015,TuZengLiHelleseth2015,WangLu2017,YuanDing2011,YuanDing2014,YuanZheng2015,ZhengChen2017,ZengZhuHu2010,ZhaHu2012,ZhaHu2016}.
Most known permutation polynomials having this form are related to $\delta$. In this paper, we find a close relationship between the two types of permutation
polynomials with the forms~(\ref{eq:ojpoly1}) and~(\ref{eq:ojpoly2}) respectively. Based on this relationship, many classes of permutation polynomials having the form~(\ref{eq:ojpoly2})
without restriction on $\delta$ are derived from known permutation polynomials of the form~(\ref{eq:ojpoly1}).

The remainder of this paper is organized as follows. In section \ref{Sec:prelim}, we introduce some preliminary results
and investigate the relationship between the two types of permutation polynomials with the forms~(\ref{eq:ojpoly1}) and~(\ref{eq:ojpoly2}) respectively.
Section~\ref{Sec:firstkind} presents four families of infinite classes of permutation trinomials with the form~(\ref{eq:ojpoly1}).
In section~\ref{Sec:secondkind}, many new classes of permutation polynomials with the form~(\ref{eq:ojpoly2}) are derived
from known permutation polynomials with the form~(\ref{eq:ojpoly1}). Finally,
concluding remark is given in section~\ref{Sec:conclud}.


\section{Preliminaries}\label{Sec:prelim}

Let $q$ be a power of a prime, and $\bF_{q}$ be a finite field with $q$ elements.
Let $k$ be a divisor of $n$. The trace function from $\bF_{q^n}$ to $\bF_{q^k}$ is defined by
\begin{equation*}
{\rm Tr}^{q^n}_{q^k}(x)= \sum_{i=0}^{n/k-1} x^{q^{ik}},
\end{equation*}
where $x\in \bF_{q^n}$. Let $d$ be a positive integer with $d\, |\, (q-1)$, and $\mu_{d}$ denote the set of $d$th
roots of unity in $\bF_{q}^*$, i.e.,
\begin{eqnarray*}
\mu_{d}=\left\{ x\in \bF_{q}^* \,|\,  x^{d}=1 \right\}.
\end{eqnarray*}

Many people investigate the permutation behavior of the polynomials with the form $x^rh(x^{s})$ for $s\,|\,(q-1)$ over $\bF_{q}$.
The following lemma is well-known which transforms the problem of proving $x^rh(x^{s})$ permutes $\bF_{q}$ into the problem of
investigating whether it permutes $\mu_{(q-1)/d}$.

\begin{lem}\label{lem:PLWZ}\cite{ParkLee2001}\cite{Wang2007}\cite{Zieve2009}
Let $d, r$ be positive integers with $d \mid (q-1)$, and let $h(x)\in \bF_q[x]$. Then $x^rh(x^{(q-1)/d})$ permutes $\bF_q$ if and only if both
\begin{enumerate}
\item[{\rm (1)}]gcd$(r,(q-1)/d)=1$ and
\item[{\rm (2)}]$x^rh(x)^{(q-1)/d}$ permutes $\mu_d:=\left\{x\in \bF_q^*\,|\, x^d=1\right\}.$
\end{enumerate}
\end{lem}

The following proposition shows a close relationship between the two types of permutation polynomials with the forms~(\ref{eq:ojpoly1}) and~(\ref{eq:ojpoly2}).

\begin{prop}\label{prop:suffice}
Let $m, k$ be integers with $0< k < m$ and $\ell = \gcd(k, m)$.
Let $\bF_{q^m}$ be a finite field and $ g(x)\in \bF_{q^m}[x]$.
For $\delta \in \bF_{q^m}$ and $c \in \bF_{q^\ell}^*$, if $h(x)=g(x)^{q^k}-g(x)+cx$
permutes $\bF_{q^m}$ then $f(x)=g(x^{q^k}-x+\delta)+cx$ permutes $\bF_{q^m}$.
\end{prop}
\pf For $\alpha\in\bF_{q^m}$, it is sufficient to show that the equation
 \begin{equation}\label{eq:fa}
  f(x)=g(x^{q^k}-x+\delta)+cx=\alpha
 \end{equation}
 has only one solution in $\bF_{q^m}$. Denote by~$y=x^{q^k}-x+\delta$.
Taking $q^k$-th power on both sides of (\ref{eq:fa}) and subtracting (\ref{eq:fa}) we have
\begin{equation*}\label{eq:ha}
h(y) = g(y)^{q^k} - g(y)+ cy = \alpha^{q^k}-\alpha + c\delta.
\end{equation*}
Since $h$ permutes $\bF_{q^m}$ we obtain $y=h^{-1}(\alpha^{q^k}-\alpha + c\delta)$. So,
the only one solution of (\ref{eq:fa}) in $\bF_{q^m}$ is
$x=c^{-1}( \alpha- g\circ h^{-1}(\alpha^{q^k}-\alpha + c\delta))$.

\begin{rem}
In Proposition~\ref{prop:suffice}, let $m=2, k=1$ and $g(x)=x^s$ and $c\in \bF_q^*$. By Proposition~\ref{prop:suffice},
if $cx-x^s+ x^{qs}$ permutes $\bF_{q^2}$ then $(x^q-x+\delta)^s+ cx$ permutes $\bF_{q^2}$.  Moreover, if there are
no restrictions on $\delta$ and $c=1$ then the inverse of Proposition~\ref{prop:suffice} also holds.
\end{rem}

\begin{prop}
Let $\bF_{q^m}$ be the degree $m$ extension of the finite field $\bF_q$.
Let $g(x)\in \bF_q[x]$ be a polynomial over $\bF_q$. For any $\delta\in \bF_{q^m}$,
$f(x)=g(x^q-x+\delta)+x$ permutes $\bF_{q^m}$ if and only if $h(x)=g(x)^q - g(x)+x$ permutes $\bF_{q^m}$.
\end{prop}
\pf By Proposition~\ref{prop:suffice} we only prove the necessity . Denote by $\varphi(x)=x^{q}-x+\delta$, $\alpha={\rm Tr}^{q^m}_q(\delta)$,
and $S_{\alpha}=\{ x\in \bF_{q^m}\, \mid\, {\rm Tr}^{q^m}_q(x)= \alpha\}$. By Theorem~2.25 in~\cite{LidlNiederreiter1983} we have that
$S_{\alpha}=\{x^q-x+\delta\, |\, x\in \bF_{q^m}\}$. For $\alpha, \alpha^\prime\in \bF_q$ with $\alpha\neq \alpha^\prime$
we have that  $S_{\alpha} \cap S_{\alpha^\prime}= \emptyset$  and $\cup_{\alpha \in \bF_q} S_{\alpha}= \bF_{q^m}$.

It is easy to verify that $\varphi \circ f (x) =h\circ \varphi(x)$ for any $x\in \bF_{q^m}$,
i.e., the following diagram is commutative,
    \[ \begin{CD}
    \bF_{q^m}                      @>f>>    \bF_{q^m} \\
    @VV \varphi V                 @VV \varphi V \\
    S_{\alpha}                           @> h>>   S_{\alpha}
    \end{CD}\]

If $f(x)$ is a bijection on $\bF_{q^m}$, then from above diagram we know that $h(x)$ is a surjective mapping from $S_{\alpha}$ to itself. So,
$h(x)$ is a bijection on $S_\alpha$ for any $\alpha\in\bF_q$, moreover $\bF_{q^m}=\cup_{\alpha \in \bF_q} S_{\alpha}$. Thus, $h(x)$ permutes $\bF_{q^m}$.
\EOP

\section{Four classes of permutation polynomials with the form $cx-x^s+ x^{sq}$ over $\bF_{q^2}$}\label{Sec:firstkind}

In this section we present four classes of permutation polynomials with the form $cx-x^s+ x^{sq}$
for some proper index $s$ and parameter $c\in \bF_{q^2}^*$.

\begin{thm}\label{thm:31}
Let $s=\frac{3q^2+2q-1}{4}$ and $c\in \bF_{q^2}^*$. The polynomial $f(x)=cx-x^s+x^{qs}$ permutes $\bF_{q^2}$ in each of the following cases:
\begin{enumerate}
\item[{\rm (1)}] $q\equiv 1 \pmod 8$ and ${(-\frac{2}{c})}^{\frac{q+1}{2}}=1$;
\item[{\rm (2)}] $q\equiv 5 \pmod8$ and ${(\frac{2}{c})}^{\frac{q+1}{2}}=1$.
   \end{enumerate}
\end{thm}

\pf We only prove the case~(1). The proof of the case~(2) is similar and the details are omitted.

Since $q\equiv 1 \pmod8$ we know that $4\, |\, (3q+5)$. The polynomial $f(x)$ is rewritten as
\[ f(x) = cx - x^{\frac{3q+5}{4}(q-1) +1} + x^{\left( \frac{3q+5}{4}q+1\right)(q-1)+1}.\]
Let $u=\frac{3q+5}{4}$. By Lemma~\ref{lem:PLWZ}, $f(x)$ permutes $\bF_{q^2}$ if and only if
  \[ g(x)=x{(c+x^{1-u}-x^u)}^{q-1} \]
  permutes $\mu_{q+1}$.

Let $\mu_{\frac{q+1}{2}}$ be the cyclic subgroup of $\mu_{q+1}$ with order $\frac{q+1}{2}$. It is easy to verify
that $\mu_{\frac{q+1}{2}}$ is exactly all square elements of $\mu_{q+1}$. Since $q\equiv 1 \pmod8$ we have that
$-1$ is non-square element of $\mu_{q+1}$. Then $-\mu_{\frac{q+1}{2}}$ is the set of all non-square elements of $\mu_{q+1}$.
Next we show that $g(x)$ permutes $\mu_{\frac{q+1}{2}}$ and $-\mu_{\frac{q+1}{2}}$ respectively.

Since $q\equiv 1 \pmod8$ we have that squaring is a bijection on $\mu_{\frac{q+1}{2}}$.
For $x\in \mu_{\frac{q+1}{2}}$, there exists an unique $y\in \mu_{\frac{q+1}{2}}$ such that $x=y^2$. Then
\[ x^u= (y^2)^u=y^{\frac{3q+5}{2}}=y,   \quad  x^{1-u} = x\cdot x^{-u} = y^2 \cdot y^{-1}=y .\]
In this case, $g(x) = c^{q-1}x$, and it permutes $\mu_{\frac{q+1}{2}}$.

For $x\in - \mu_{\frac{q+1}{2}}$, there exists an unique $y\in \mu_{\frac{q+1}{2}}$ such that $x=-y^2$
with the same reason above. Due to ${(-\frac{2}{c})}^{\frac{q+1}{2}}=1$, i.e., $\frac{c}{2} \in -\mu_\frac{q+1}{2}$
and $y\in \mu_{\frac{q+1}{2}}$ we have that
$c-2y\neq 0$. So,
\begin{align*}
    g(x)=& g(-y^2)= -y^2{(c+{(-1)}^{1-u}y-{(-1)}^uy)}^{q-1} \\
        =& -y^2\frac{{(c-2y)}^q}{c-2y} =  -y^2 \frac{c^q-2y^{-1}}{c-2y}\\
        =&\frac{2}{c}y.
\end{align*}
Since  $\frac{c}{2}\in -\mu_\frac{q+1}{2}$ and $y\in \mu_{\frac{q+1}{2}}$ we have $\frac{2}{c}y \in -\mu_\frac{q+1}{2}$.
So, $g(x)$ permutes $-\mu_\frac{q+1}{2}$. Combining these two cases, we have that $g(x)$ permutes $\mu_{q+1}$.
\EOP

Similarly, we obtain another family of infinite class of permutation polynomials over $\bF_{q^2}$ as follows.

\begin{thm}\label{thm:32}
Let $s=\frac{(q+1)^2}{4}$ and $c\in \bF_{q^2}^*$. The polynomial $f(x)=cx-x^s +x^{qs}$ permutes $\bF_{q^2}$ in each of the following cases:
\begin{enumerate}
\item[{\rm (1)}] $q\equiv 5 \pmod8$ and ${(-\frac{2}{c})}^{\frac{q+1}{2}}=1$;
\item[{\rm (2)}] $q\equiv 1 \pmod8$ and ${(\frac{2}{c})}^{\frac{q+1}{2}}=1$.
\end{enumerate}
\end{thm}

By direct verification and simplification we get a family of infinite class of permutation polynomials over $\bF_{q^2}$
as follows.

\begin{thm}\label{thm:33}
 Let $q$ be a power of a prime with~$q\equiv 1\pmod3$, and $s=\frac{q^2+q+1}{3}$. Then $f(x)=x-x^s +x^{qs}$ is a permutation polynomial over $\bF_{q^2}$.
\end{thm}
\pf
Since $q\equiv 1 \pmod 3$ we know that $3\, |\, (q+2)$. The polynomial $f(x)$ is rewritten as
\[ f(x) = x - x^{\frac{q+2}{3}(q-1) +1} + x^{\left( \frac{q(q+2)}{3}+1\right)(q-1)+1}.\]
Let $u=\frac{q+2}{3}$. By Lemma~\ref{lem:PLWZ}, $f(x)$ permutes $\bF_{q^2}$ if and only if
  \[ g(x)=x{(1+x^{1-u}-x^u)}^{q-1} \]
permutes $\mu_{q+1}$.

First, we show that $1+x^{1-u}-x^u=0$ has no roots in $\mu_{q+1}$. Otherwise, assume that
there is $x\in \mu_{q+1}$ satisfying this equation, i.e.,
\begin{equation}\label{eq:thm31}
1+x^{1-u}-x^u=0, \quad x^{q+1}=1.
\end{equation}
These imply $x^{2u-1}-x^{u-1}=x^{q+1}$, i.e., $x^{2u-2}-x^{u-2}=x^q$. This leads to $x^{2-2u}-x^{2-u}=x$, then $x^{1-2u}-x^{1-u}=1$,
i.e., $x^{2u-1}+x^{u}=1$. This equality together with (\ref{eq:thm31}) implies $x^{u-1}+x^u=0$, i.e., $x=-1$.
On the other hand, it is easy to verify that $x=-1$ does not satisfy the first equation in (\ref{eq:thm31}) since $3 \nmid q$.
So, we get a contradiction.

For any $x\in \mu_{q+1}$, we have $x^{3u}=x$ since $u=\frac{q+2}{3}$, and
\begin{align*}
g(x)=& x\frac{1+x^{u-1}-x^{-u}}{1+x^{1-u}-x^u}=\frac{x+x^u-x^{1-u}}{1+x^{1-u}-x^u}=\frac{x^u+ x^{3u}-x^{2u}}{1+x^{2u}-x^u}=x^u.
\end{align*}
So, $g(x)$ permutes $\mu_{q+1}$ since ${\gcd}(u,q+1)=1$.
\EOP

The main technique in the proof of above theorems is provided in Lemma~\ref{lem:PLWZ}. This method doesn't work for
the following theorem. Using the method provided by Dobbertin~\cite{Dobbertin2002} and Kyureghyan and Zieve~\cite{KyureghyanZieve2016}
we present a family of infinite class of permutation polynomials over $\bF_{q^4}$. To this end, we first give two preliminary lemmas.

\begin{lem}\label{lem:case21}
There exists $x\in \bF_{q^4}^*$ such that $x^{2q^2}+x^{q^2+1}+x^2=0$ if and only if $3 \mid q$ and $x^{q^2-1}=1$.
\end{lem}
\pf
Assume that $x\in \bF_{q^4}^*$ satisfies
\begin{equation}
x^{2q^2}+x^{q^2+1}+x^2=0.
\end{equation}
Let $y=x^{q^2-1}$, then $y\in \mu_{q^2+1}$ and $y^2+y+1=0$. It follows that $y^3=1$. Since ${\rm gcd}(3,q^2+1)=1$,
we must have $y=1$ and $3|q$. Conversely, it is obvious.
\EOP

Similarly, we have the following lemma.
\begin{lem}\label{lem:case22}
There exists $x\in \bF_{q^4}^*$ such that $x^{2q^2}-x^{q^2+1}+x^2=0$ if and only if $3 \mid q$ and $x^{q^2-1}=-1$.
\end{lem}


\begin{thm}\label{thm:34}
Let $q$ be a power of an odd prime and $s=q^3+q^2-q$. Then the polynomial $f(x)=x-x^s +x^{q^2 s}$ permutes $\bF_{q^4}$.
\end{thm}
\pf We prove that $f(x)=\alpha$ has only one root in $\bF_{q^4}$ for any $\alpha\in \bF_{q^4}$. To this end,
we discuss the proof according to the following four situations.

{\bf Case I:} $\alpha=0$ and $f(x)=0$. It is clear that this equation has a solution $x=0$. Next we show that there is no
$x \in \bF_{q^4}^*$ satisfying $f(x)=0$. Otherwise, we have
  \[ 1-x^{q^3+q^2-q-1}+x^{-q^3+q}=0 .\]
Denote by $y=x^{q^2-1}$, then $y\in \mu_{q^2+1}$. From above equation we get
\begin{equation}\label{eq:thm41}
    y^{q+1}-y^{-q}=1.
\end{equation}
Calculation of ${(\ref{eq:thm41})}^q-{(\ref{eq:thm41})}^{q^3}$ together with $y^{q^2}=y^{-1}$, we obtain
\[ \begin{split}
0=&y^{q^2+q}-y^{-q^2}- y^{q^3+1}+y^{-1}\\
 =&y^{q-1} -y-y^{-q+1} + y^{-1}\\
 =&(y^{q-1}-y)(y^{-q}+1).
\end{split} \]
It follows that $y^{q-2}=1$ since $y\neq -1$. This fact together with $y^{q^2+1}=1$ implies that
\[ y^{{\rm gcd}(q-2,q^2+1)}=y^{{\rm gcd}(q-2,5)}=1 .\]
So, we have $y^5=1$. On the other hand, substituting $y^{q-2}=1$ into (\ref{eq:thm41}) we get
$y^5=y^2+1$. This together with $y^5=1$ implies that $y=0$. This is a contradiction. So, $f(x)=0$
has only one solution $x=0$ in $\bF_{q^4}$.

To discuss case of $\alpha\neq 0$, we define two sets as follows,
\[ S_{\pm} = \left\{ x\in \bF_{q^4}^* \mid x^{2q^2} \pm x^{q^2+1}+x^2=0 \right\} .\]

{\bf Case II:} $\alpha \in S_{+}$ and $f(x)=\alpha$. By Lemma~\ref{lem:case21}, we have $3 \mid q$ and $\alpha^{q^2}-\alpha=0$.
In this case we have
\[\begin{split}
0=&\alpha^{q^2}-\alpha=f(x)^{q^2} -f(x) \\
 =&x^{q^2}-x^{-q^3+q+1}+x^{q^3+q^2-q}-(x-x^{q^3+q^2-q}+x^{-q^3+q+1})\\
 =& x^{q^2}-x-x^{q^3+q^2-q}+x^{-q^3+q+1}\\
 =&x^{-q^3-q}{(x-x^{q^2})}^q(x^{q^2+q^3}+x^{1+q}).
\end{split}
 \]
If $x-x^{q^2}=0$ then $f(x)=x=\alpha$. If $x^{q^2+q^3}=-x^{1+q}$ then $x^{(q+1)(q^2-1)}=-1$.
Since $x\in \bF_{q^4}^*$, we have $x^{{\gcd}(2(q+1)(q^2-1),q^4-1)}=x^{2(q+1)}=1$. It implies that $x^{(1+q)(q^2-1)}=1$.
This leads to a contradiction. Thus $f(x)=\alpha$ has only one root $x=\alpha$.

{\bf Case III:} $\alpha \in S_{-}$ and $f(x)=\alpha$. By similar proof to Case II, we know that $f(x)=\alpha$ has only one
solution $x=\alpha$.

Denote by $T=\bF_{q^4}^*\setminus (S_{+}\cup S_{-})$. From the proving process of Cases II and III, we know that $f(x)$ is
a permutation over $S_{+}$ and $S_{-}$ respectively. Moreover, $f(T)\subseteq T$. It remains to show that $f(x)$ is also a
permutation over the set $T$.

{\bf Case IV:} $\alpha\in T$. We will show that $f(x)=\alpha$ has only one $x\in T$ satisfying this equation. Denote by
$$ y=x^q,\,\, z=x^{q^2},\,\, w=x^{q^3},\,\, \beta=\alpha^q,\,\, \gamma=\alpha^{q^2},\,\, \delta=\alpha^{q^3}.$$
The equation $f(x)=\alpha$ is reduced to
\begin{equation}\label{eq:thm42}
  x+{\frac {xy}{w}}-{\frac {zw}{y}}=\alpha.
\end{equation}
Taking $q$th, $q^2$th and $q^3$th power on both sides of~(\ref{eq:thm42}) respectively, we obtain
\begin{equation}\label{eq:thm43}
  y+{\frac {yz}{x}}-{\frac {wx}{z}}=\beta,
\end{equation}
\begin{equation}\label{eq:thm44}
  z+{\frac {zw}{y}}-{\frac {yx}{w}}=\gamma,
\end{equation}
\begin{equation}\label{eq:thm45}
  w+{\frac {wx}{z}}-{\frac {yz}{x}}=\delta,
\end{equation}
respectively. Adding (\ref{eq:thm42}) and (\ref{eq:thm44}) we have $z=\alpha+\gamma-x$. Similarly, we get $w=\beta+\delta-y$ from (\ref{eq:thm43}) and (\ref{eq:thm45}).
Substituting the two equalities into (\ref{eq:thm42}) and (\ref{eq:thm43}) respectively, we obtain
\begin{equation}\label{eq:thm46}
\left\{ \begin{split}
&( -\gamma+x ) {y}^{2}+ \left( ( \alpha+2\,\gamma-x ) (\beta+\delta) \right) y+( -\alpha-\gamma+x ) (\beta+\delta)^2=0 \\
&\left( {x}^{2}- ( \alpha+\gamma ) x+ ( \alpha+\gamma)^{2} \right) y-\delta x^{2}-\beta (\alpha+\gamma ) x=0 .
\end{split} \right.
\end{equation}
Since $x\in T$, from Lemma~\ref{lem:case21} we have
$${x}^{2}- ( \alpha+\gamma ) x+ ( \alpha+\gamma)^{2}=x^2+xz+z^2\neq 0.$$
From the second equation in (\ref{eq:thm46}) we get
\begin{equation}\label{eq:thm47}
y =\frac {x ( \beta\alpha+\beta\gamma+\delta x) }{\alpha^2+2\alpha\gamma+\gamma^{2}-(\alpha+\gamma) x+x^2} \triangleq H(x).
\end{equation}
Substituting this $y$ into the first equation in (\ref{eq:thm46}), we obtain a equality $A(x)=0$,
where $A(x)\in\bF_{q^4}^*[\alpha,\beta,\gamma,\delta][x]$ and its degree on $x$ is $5$.

Suppose that $f(x)=\alpha$ have another root $X \in T $ different from $x$, then $A(X)=0$.
Replacing $\alpha,\beta,\gamma,\delta$ in coefficients of $A(X)$ with $x,y,z,w$ respectively,
and multiplying $wx^2yz^2$ on both sides of $A(X)=0$, we get
\begin{equation}\label{eq:bc=0}
 (X-x) B(X) C(X) = 0 ,
\end{equation}
where
\[ \begin{split}
B(X)=&yw(x^{2}+xz+z^{2})X^{2}+wxz ( z+x)^{2}( w+y) \\
-&(z + x)( {w}^{2}xz+w{x}^{2}y+wxyz+wy{z}^{2}-x{y}^{2}z)X,\\
C(X)=&(w^{2}x^{2}-wxyz+y^{2}{z}^{2})X^2+yz ( z+x)^3 ( w+y)\\
    +&(z+x)( w^{2}xz+wx^{2}y+wxyz-wyz^{2}-xy^{2}z-2y^{2}z^{2})X.
\end{split} \]
Hence $B(X)=0$ or $C(X)=0$. It has been verified that $yw(x^{2}+xz+z^{2})\neq 0$ above.
Assume that $w^{2}x^{2}-wxyz+y^{2}{z}^{2}= 0$. Then
\begin{equation}\label{eq:lcofc}
0 = w^{2}x^{2}-wxyz+y^{2}{z}^{2} = (yz)^{2q^2}-(yz)^{q^2+1} + (yz)^2 .
\end{equation}
By Lemma~\ref{lem:case22} we have $(yz)^{q^2-1}=-1$, i.e., $y^{(q^2-1)(q+1)}=-1$. Since
$\gcd(2(q^2-1)(q+1), q^4-1)=2(q^2-1)$, we have $y^{2(q^2-1)}=1$. This is a contradiction
with $y^{(q^2-1)(q+1)}=-1$. So, the degree of $B(X)$ and $C(X)$ on $X$ is $2$, respectively.

(i) When $B(X)=0$ in (\ref{eq:bc=0}). Taking $q$th power on both sides of this equality we get $B^\prime(x^q) =0$, where
the coefficients of $B^\prime(X)$ are $q$th power of corresponding coefficients of $B(X)$. Replacing $Y=X^q$
with $H(X)$ in (\ref{eq:thm47}) we obtain
\begin{equation}\label{eq:cd=0}
 0 = B^\prime(H(X)) = C(X) D(X),
\end{equation}
where
\[
\begin{split}
D(X)=&( w^{2}x^{2}+{w}^{2}xz+zxwy+x{y}^{2}z+{y}^{2}{z}^{2})X^{2}+wx^{2}(z+x)^{2}( w+y)\\
    -&( z+x)( 2w^{2}{x}^{2}+{w}^{2}xz+w{x}^{2}y+zxwy-wy{z}^{2}+x{y}^{2}z) X.
 \end{split}
\]

If $C(X)=0$ from (\ref{eq:cd=0}) then $X$ is a common root of $B(X)$ and $C(X)$. Then the resultant of these two polynomials should be $0$. So,
\begin{equation}\label{eq:thm48}
 \begin{split}
0=Res(B, C, X)&=zx(z+x)^{4}(w+y)^2(w^{2}{x}^{2}-zxwy+y^{2}z^{2}) \\
              &\times(w^2xz+w^2z^2+wxyz+x^2y^2+xy^2z)^{q+1}.
\end{split}
\end{equation}
It has been verified that $w^{2}{x}^{2}-zxwy+y^{2}z^{2}\neq 0$ above. Moreover, we have that $(z+x)(w+y)\neq 0$.
In fact, if $x+z=0$ or $w+y=0$ we can derive that $B(X)=D(X)=x^2y^2 X^2 =0$.
From (\ref{eq:thm48}) we have
\begin{equation}\label{eq:thm49}
w^2xz+w^2z^2+wxyz+x^2y^2+xy^2z=0.
\end{equation}
Combining (\ref{eq:thm49}) and $B(X)=0$ we obtain that
\[ yw(x^2+xz+z^2)(X-x)^2 = 0 .\]
This is a contradiction with that $x^2+xz+z^2\neq 0$ and $X\neq x$.

So, from (\ref{eq:cd=0}) we must have $D(X)=0$. If the leading coefficient of $D(X)$ is nonzero,
i.e., $w^{2}x^{2}+{w}^{2}xz+zxwy+x{y}^{2}z+{y}^{2}{z}^{2}\neq 0$ then
\[0 = Res(B, D, X) = xwyz(z+x)^{6}(w+y)^{2}\left( w^{2}x^{2}-zxwy+y^{2}z^{2}\right)^{q+1} .\]
This is impossible from the discussion in previous paragraph. Thus, the leading coefficient of $D(X)$ is zero, i.e.,
\begin{equation}\label{eq:lcofd}
L(x) = {w}^{2}{x}^{2}+{w}^{2}xz+zxwy+x{y}^{2}z+{y}^{2}{z}^{2} =0 .
\end{equation}
The fact that $w{x}^{2}( z+x) ^{2}( w+y)\neq 0$ implies that the degree of $D(X)$ is 1.
The resultant of $B(X)$ and $D(X)$ is zero. Then
\begin{equation}\label{eq:resbd}
0= Res(B(X), D(X), X) = (z+x)^4xw(w+y)U(x),
\end{equation}
where
\[ \begin{split}
U(x)=&-yw^{3}x^{5}+(2zw^{4}+yzw^{3}+2y^{2}zw^{2}x^{4})x^{4}+(3z^{2}w^{4}+3yz^{2}w^{3}\\
    +&6y^{2}z^{2}w^{2}+2y^{3}z^{2}w+y^{4}z^{2})x^{3}+(z^{3}w^{4}-2 yz^{3}w^{3}+y^{2}z^{3}w^{2}\\
    +&y^{4}z^{3})x^{2}-(2yz^{4}w^{3}+y^{2}z^{4}w^{2}+2y^{3}z^{4}w)x+y^{2}z^{5}w^{2}.
\end{split}
\]
From (\ref{eq:resbd}) we have $U(x)=0$. By calculation of the resultant of $L(x)$ and $U(x)$ we get
\begin{equation}\label{eq:reslu}
0=Res(L(x), U(x), x) = ( {w}^{2}+{y}^{2})^{2} ( w+y)^{6}{w}^{4}{z}^{10}{y}^{4} .
\end{equation}
In this case, it is verified that $y+w\neq 0$ above. From (\ref{eq:reslu}) we have $w^2=-y^2$, i.e.,
$y^{2(q^2-1)}=-1$ and $y^{4(q^2-1)}=1$. Then $y^{\gcd(4(q^2-1),q^4-1)}=y^{2(q^2-1)}=1$.
This is a contradiction. Therefore, we prove that $B(X)$ can not be zero.

(ii) Similarly, we can prove that $C(x)$ in (\ref{eq:bc=0}) is not zero. Thus, $f(x)=\alpha$ has only one
root in $T$, and $f(x)$ permutes $T$.
 \EOP


\section{Permutation polynomials with the form $(x^q-x+\delta)^s +cx$ over~$\bF_{q^2}$ }\label{Sec:secondkind}

In this section we give many classes of permutation polynomials having the form $(x^q-x+\delta)^s +cx$
without restriction on $\delta$ over $\bF_{q^2}$. These permutation polynomials are derived from known
permutation trinomials with the form~(\ref{eq:ojpoly1}) by using Proposition~\ref{prop:suffice}.

\subsection{Permutation polynomials over finite fields with odd characteristic }

First, four families of infinite classes of permutation polynomials of the form $(x^q-x+\delta)^s +cx$ over
$\bF_{q^2}$ with odd characteristic are derived from permutation trinomials introduced in Section~\ref{Sec:firstkind}.

\begin{thm}
Let $\delta\in \bF_{q^2}$ and $c\in \bF_{q}^*$. The polynomial
$$f(x)={(x^{q}-x+\delta)}^{\frac{3q^2+2q-1}{4}}+cx$$
is a permutation over $\bF_{q^2}$ in each of the following cases:
\begin{enumerate}
\item[{\rm (1)}] $q\equiv 1 \pmod8$ and $c=-2$;
\item[{\rm (2)}] $q\equiv 5 \pmod8$ and $c=2$.
\end{enumerate}
\end{thm}
\pf Let $s=\frac{3q^2+2q-1}{4}$ and $g(x)=x^s$. From Theorem~\ref{thm:31} we know that $h(x)=cx+g(x)^s-g(x)^{qs}$
permutes $\bF_{q^2}$ under the case of $q\equiv 1 \pmod 8$ and $(-2/c)^{\frac{q+1}{2}}=1$ or the case of
$q\equiv 5 \pmod 8$ and $(2/c)^{\frac{q+1}{2}}=1$ respectively. These are exact cases listed in (1) and (2) respectively
since $c\in \bF_q$. By Proposition~\ref{prop:suffice} we obtain that $f(x)$ permutes
$\bF_{q^2}$ under one of conditions listed above.
\EOP

Similarly, one easily verifies the following three theorems.
\begin{thm}
Let $\delta\in \bF_{q^2}$ and $c\in \bF_{q}^*$. The polynomial
$$f(x)={(x^{q}-x+\delta)}^{\frac{{(q+1)}^2}{4}}+cx$$ is a permutation over $\bF_{q^2}$ in each of the following cases:
\begin{enumerate}
\item[{\rm (1)}] $q\equiv 1 \pmod8$ and $c=2$;
\item[{\rm (2)}] $q\equiv 5 \pmod8$ and $c=-2$.
\end{enumerate}
\end{thm}

\begin{thm}
Let $q$ be a power of a prime with $q\equiv 1 \pmod{3}$. For $\delta\in \bF_{q^2}$, the polynomial
$$ f(x)={(x^{q}-x+\delta)}^{\frac{q^2+q+1}{3}}+x $$
is a permutation over $\bF_{q^2}$.
\end{thm}

\begin{thm}
Let $q$ be a power of an odd prime. For any $\delta\in \bF_{q^2}$, the polynomial
$$f(x)={(x^{q}-x+\delta)}^{q^3+q^2-q}+x$$
is a permutation over $\bF_{q^4}$.
\end{thm}

\subsection{Permutation polynomials over finite fields with even characteristic}

By Proposition~\ref{prop:suffice} many classes of permutation polynomials having the form
$(x^q+x+\delta)^s +cx$ without restriction on $\delta$ are derived from known permutation trinomials
having the form $cx +x^s + x^{qs}$ over $\bF_{q^2}$ with even characteristic. To this end,
we first recall known permutation trinomials with the form $cx +x^s + x^{qs}$ over finite fields with
even characteristic in the following lemmas.

\begin{lem}\label{lem:qu2017}\cite{LiQuChenLi2017}
Let $k$ be a positive integer and $q=2^k$. The trinomials $f(x)=cx+x^s + x^{qs}$ are permutation polynomials over $\bF_{q^2}$
in each of the following cases.
\begin{enumerate}
\item[{\rm(1)}]$s=2q-1$. The positive integer $k$ and $c\in \bF_{q^2}$ satisfy one of the following conditions:
\begin{enumerate}
\item[{\rm i)}] $k$ is even and $c=1$;
\item[{\rm ii)}] $k$ is odd and $c^3=1$.
\end{enumerate}
\item[{\rm(2)}]$s=\frac{(3q-2)(q^2+q+1)}{3}$, and $k$ is even and~$c\in \bF_{q^2}$ satisfies~$c^3=1$.
\item[{\rm(3)}]$s=\frac{q+4}{5}$, and $k$ is odd and~$c\in \bF_{q^2}$ satisfies~$c^3=1$.
\item[{\rm(4)}]$s=\frac{3q+1}{4}$, and $c\in \bF_{q}$ satisfies that $x^3+x+c=0$ having no solution in $\bF_q$.
\item[{\rm(5)}]$s=\frac{q+6}{7}$ and $c=1$.
\item[{\rm(6)}]$s=\frac{q^2+3q+2}{6}$, and $k$ is odd and $c\in \bF_{q^2}$ satisfies $c^{\frac{q+1}{3}}=1$.
\item[{\rm(7)}]$s=\frac{q^2-2q+4}{3}$, and $k$ is even and $c=1$.
\item[{\rm(8)}]$s=\frac{Q^3+Q^2-Q+1}{2}$ and $Q=2^{\frac{k}{2}}$ for an even $k$, and $c\in \bF_Q^*$.
    \end{enumerate}
\end{lem}

\begin{lem}\label{lem:linian}\cite{LiHelleseth2016}\cite{BartoliQuoos2017}
Let $k$ be a positive integer and $q=2^k$. The trinomials $f(x)=cx+x^s + x^{qs}$ are permutation polynomials over $\bF_{q^2}$
in each of the following cases.
\begin{enumerate}
\item [{\rm (1)}]$s=\frac{-1}{2^{k^\prime}-1}(2^k-1)+1$, where ${k^\prime}$ is a positive integer with $\gcd(2^{k^\prime}-1,2^k+1)=1$ and $c\in \bF_{2^{k^\prime}}^*\cap \bF_{q}$.
\item [{\rm (2)}]$s=\frac{1}{2^{k^\prime}+1}(2^k-1)+1$, where ${k^\prime}$ is a positive integer with $\gcd(2^{k^\prime}+1,2^k+1)=1$ and   $c\in \bF_{2^{k^\prime}}\cap \bF_{q}$ .
\end{enumerate}

\end{lem}

\begin{rem}
It is easy to see that each $s$ in Lemmas~\ref{lem:qu2017} and \ref{lem:linian} can be rewritten as the form
$s=i(q-1)+1$, here $i$ is interpreted as modulo $q+1$ when it is negative or proper fraction.
\end{rem}

By Proposition~\ref{prop:suffice}, the coefficient $c$ of $x$ in polynomials of the form (\ref{eq:ojpoly2})
should be in $\bF_{q}^*$, and the permutation polynomials in the following theorem are derived directly
from permutation trinomials in Lemmas~\ref{lem:qu2017} and \ref{lem:linian}.

\begin{thm}\label{genaralizition}
Let $k$ be a positive integer and $q=2^k$. For $\delta\in \bF_{q^2}$, $f(x)={(x^q+x+\delta)^s+cx}$
permutes~$\bF_{q^2}$ in each of the following cases:
\begin{enumerate}
\item[{\rm(1)}]$s=2q-1$ and $c=1$.
\item[{\rm(2)}]$s=\frac{(3q-2)(q^2+q+1)}{3}$ and $k$ is even, and $c\in \bF_q$ satisfies $c^3=1$.
\item[{\rm(3)}]$s=\frac{q+4}{5}$, $c=1$ and~$k$ is odd.
\item[{\rm(4)}]$s=\frac{3q+1}{4}$, and $c\in \bF_{q}$ satisfies that $x^3+x+c=0$ having no solution in $\bF_q$.
\item[{\rm(5)}]$s=\frac{q+6}{7}$ and $c=1$.
\item[{\rm(6)}]$s=\frac{q^2+3q+2}{6}$, $c=1$ and $k$ is odd.
\item[{\rm(7)}]$s=\frac{q^2-2q+4}{3}$, $c=1$ and $k$ is even.
\item[{\rm(8)}]$s=\frac{Q^3+Q^2-Q+1}{2}$ and $Q=2^{\frac{k}{2}}$ for an even $k$, and $c\in \bF_{Q}^*$.
\item[{\rm(9)}]$s=\frac{-1}{2^{k'}-1}(2^k-1)+1$, and $k'$ is a positive integer with $\gcd(2^{k'}-1,2^k+1)=1$, and $c\in \bF_{2^{k\prime}}\cap \bF_q$.
\item[{\rm(10)}]$s=\frac{1}{2^{k'}+1}(2^k-1)+1$, and $k'$ is a positive integer with $\gcd(2^{k'}+1,2^k+1)=1$, and  $c\in \bF_{2^{k^\prime}}\cap \bF_q$.
\end{enumerate}
\end{thm}

\begin{rem}
The permutation polynomial in case {\rm (1)} has been proposed in Theorem~1 of~\cite{TuZengJiang2015} and Theorem~3.2 of~\cite{WangLu2017}. 
The permutation polynomial in case~{\rm (3)} has been constructed in~\cite{GuptaSharma2018}. The permutation polynomial in case~{\rm (4)}
has been constructed in Theorem~2.1 of~\cite{ZhaHu2016} and Theorem~3.1 of~\cite{WangLu2017}.
Even though the representation of permutation polynomials having the form {\rm (\ref{eq:ojpoly2})} in those papers may be different,
they have the same corresponding trinomials of the form {\rm (\ref{eq:ojpoly1})} if $s$ is written as $i(q-1)+1$ and $i$ is interpreted 
as modulo $q+1$ when it is negative or proper fraction. As far as we know, Table~1 presents all permutation polynomials 
having the form $(x^q+x+\delta)^s + cx$ without restriction on $\delta$ over $\bF_{q^2}$
with even characteristic, where $\Omega=\{c\in \bF_{q} \mid x^3+x+c=0 \text{  has no roots in  } \bF_q \}$.
\end{rem}

\begin{table}[htbp]\label{table1}
\renewcommand\arraystretch{1.25}
\caption{\rm  PPs of the form $(x^{2^k}+x+\delta)^s +cx$ without restriction on $\delta$ over $\bF_{2^{2k}}$ }
\begin{center}
{ \footnotesize
\begin{tabular}{cccccc}\hline
  No. & $k$  & $s=i(2^k-1)+1$ & $c$  &  Reference \\
  \hline
  1 & $k$ is even  & $i=2$ or$-1$        & $c=1$  &  \cite{TuZengJiang2015}\cite{WangLu2017}\\
  \hline
  2 & all $k$      & $i=0$ or $1$        &  $c=1$  & \cite{WangLu2017}\cite{YuanDing2011}\\
  \hline
  3 & all $k$      & $i=\frac{1}{2}$     &  $c=1$  & \cite{WangLu2017} \\
  \hline
  4 & $k$ is even  & $i=\frac{1}{3}$ or $\frac{2}{3}$  &  $c=1$  &  \cite{XuCaoXu2016}\\
   \hline
  5 & $k$ is odd   &   $i=\frac{1}{5}$  or $\frac{4}{5}$         &  $c=1$  &\cite{GuptaSharma2018}\\
  \hline
  6 &  a positive integer    & $i=\frac{1}{4}$ or$\frac{3}{4}$    & $c\in \Omega$    &  \cite{ZhaHu2016}\cite{WangLu2017}  \\
  \hline
  7 & $k$ is even  & $i=\frac{1}{2^k-2}$ or $\frac{-4}{2^k-2}$  &  $c^3=1$  & Theorem~\ref{genaralizition} \\
  \hline
  8 &  a positive integer   & $i=\frac{1}{7}$ or$\frac{6}{7}$          &$c=1$  &  Theorem~\ref{genaralizition}\\
  \hline
  9 & $k$ is odd          & $i=\frac{2^k+4}{6}$ or$\frac{2-2^k}{6}$          & $c^\frac{q+1}{3}=1$ & Theorem~\ref{genaralizition} \\
  \hline
  10 & $k$ is even        & $i=\frac{2^k-1}{3}$ or $\frac{4-2^k}{3}$         &$c=1$& Theorem~\ref{genaralizition}\\
  \hline
  11 & $k$ is even         & $i=\frac{2^{k/2}+1}{2}$ or $\frac{1-2^{k/2}}{2}$    &  $c\in\bF_{2^{k/2}}^*$  & Theorem~\ref{genaralizition}  \\
  \hline
  12 & $\gcd(2^{k'}-1,2^k+1)=1$          & $i=\frac{-1}{2^{k'}-1}$ or $\frac{2^{k'}}{2^{k'}-1}$        &  $c\in\bF_{2^k}^*\bigcap \bF_{2^{k'}}$  & Theorem~\ref{genaralizition} \\
  \hline
  13 & $\gcd(2^{k'}+1,2^k+1)=1$         & $i=\frac{1}{2^{k'}+1}$ or $\frac{2^{k'}}{2^{k'}+1} $         &  $c\in\bF_{2^k}^*\bigcap \bF_{2^{k'}}$  &  Theorem~\ref{genaralizition} \\
  \hline
\end{tabular}}
\end{center}
\end{table}

\section{Concluding remark}\label{Sec:conclud}

In this paper, we proposed four families of infinite classes of permutation trinomials
with the form~(\ref{eq:ojpoly1}). From known permutation trinomials with the form~(\ref{eq:ojpoly1}),
many classes of permutation polynomials with the form~(\ref{eq:ojpoly2}) are derived.
As part of the future work, observe that the relationship between the two types of polynomials can be
generalized to the case of multiple monomials. It would be interesting to find more permutation polynomials with the
form~(\ref{eq:ojpoly1}) or (\ref{eq:ojpoly2}) by using this relation.

\end{document}